# A Comparative Analysis of Machine Learning Algorithms for Intrusion Detection in Edge-Enabled IoT Networks

Poornima Mahadevappa
*School of Computer Science and Engineering*
*Taylor's University*
Subang Jaya, Malaysia
poornimamahadevappa@sd.taylors.edu.my

Syeda Mariam Muzammal
*School of Computer Science and Engineering*
*Taylor's University*
Subang Jaya, Malaysia
syedamariammuzammal@sd.taylors.edu.my

Raja Kumar Murugesan*
*School of Computer Science and Engineering*
*Taylor's University*
Subang Jaya, Malaysia
rajakumar.murugesan@taylors.edu.my

*Abstract*— A significant increase in the number of interconnected devices and data communication through wireless networks has given rise to various threats, risks and security concerns. Internet of Things (IoT) applications is deployed in almost every field of daily life, including sensitive environments. The edge computing paradigm has complemented IoT applications by moving the computational processing near the data sources. Among various security models, Machine Learning (ML) based intrusion detection is the most conceivable defense mechanism to combat the anomalous behavior in edge-enabled IoT networks. The ML algorithms are used to classify the network traffic into normal and malicious attacks. Intrusion detection is one of the challenging issues in the area of network security. The research community has proposed many intrusion detection systems. However, the challenges involved in selecting suitable algorithm(s) to provide security in edge-enabled IoT networks exist. In this paper, a comparative analysis of conventional machine learning classification algorithms has been performed to categorize the network traffic on NSL-KDD dataset using Jupyter on Pycharm tool. It can be observed that Multi-Layer Perception (MLP) has dependencies between input and output and relies more on network configuration for intrusion detection. Therefore, MLP can be more appropriate for edge-based IoT networks with a better training time of 1.2 seconds and testing accuracy of 79%.

## I. INTRODUCTION

Cybersecurity has emerged as a crucial research aspect in Internet of Things (IoT) due to the increasing cyber threats. The purpose of IoT security is to reduce risks for critical applications and sensitive data. IoT has been a fundamental component for many applications such as smart cities, smart grids, smart manufacturing, healthcare, and many more. As the number of IoT devices increases, potential security exposures are also growing exponentially. The pervasive nature of IoT devices and the associated applications shows the importance of addressing these security threats [1]. For instance, in IoT-enabled smart home appliances connected to the local network, the attackers can hack sensitive data to alter or interrupt the workflow. Similarly, another instance of extracting personal information through a patient's pacemaker and changing the device's behavior [2]. Currently, there are many innovative techniques to mitigate cybersecurity issues, such as deploying encryption technologies, frequent device updates, develop documented guidelines, and many more. However, by 2025 more than 30 billion devices are expected globally, it can be challenging to provide cyber security to these abundant devices [3].

Edge computing is an emerging decentralized computing paradigm that aims to enhance security, prevent data theft, minimize data flow and improve the efficiency of IoT applications. Many IoT applications have benefited from edge computing, like reducing network bandwidth, computing time, and the frequency of two-way communication between devices and the cloud. Since many IoT devices are resource-constrained, the attackers can make them vulnerable to security threats. But the edge nodes in the edge computing architecture can perform sophisticated security functions and secure IoT devices [4]. Real-time services, transient storage, data dissemination, and decentralized computation are some of the security solutions which edge computing can provide to IoT devices. Edge nodes provide real-time services like intrusion detection and identity authentication by performing computing near the data generation points. The sensitive data acquired from IoT devices is protected and shared securely by edge nodes temporarily through transient storage. Finally, encryption techniques during data dissemination and verifiable or server-aided decentralized computation are the solutions incorporated by edge computing. Among all these, intrusion detection by edge computing is an efficient solution that can identify the attacks targeting local services. Edge nodes identify any policy violations or malicious activities and prevent them from affecting the whole infrastructure. In a complex attack situation, they collaborate with the adjacent nodes or the nodes in a higher hierarchy to detect the attack. With a reliable intrusion detection system in edge-enabled applications, this cooperation of edge nodes can improve malicious attack detection success rate [5].

The area of Machine Learning (ML) has a significant interest in many domains. It is widely used in IoT security to provide defense against attacks compared to traditional methods. Integrating intrusion detection on edge nodes with ML algorithms offers a promising platform to overcome the challenges for securing IoT devices from cyberattacks. The role of ML is to use and train the algorithms to detect anomalies or any suspicious activities in the network. ML algorithms can perform two-step processes such as learning and classification steps whenever the available data is labelled. The learning step trains the model, and during the classification, the model predicts the new label of the data. This process makes the model detect any new attacks that traditional security algorithms could not detect [6].

This paper compares ML classification algorithms adopted on edge nodes to analyze the input data stream and network traffic. The study uses the NSL-KDD dataset since it contains the labelled records of simple intrusion detection networks. The labels, as normal or attack, and scores indicating the severity of attacks makes it suitable to compare the efficiency of ML classifiers. Jupyter on Pycharm tool is used to analyze the algorithms.

The organization of the rest of the paper is as follows: Section II describes the recent trends in existing literature for intrusion detection in edge-based IoT applications. Section III presents the comparative analysis of different machine learning classification algorithms. Section IV discusses the findings for the research domain under study. Section V provides the conclusion.

## II. Recent Trends in Edge-based Intrusion Detection

IoT comprises various interconnected devices to gather, process, refine, and exchange data over the internet. These devices have respective identities or IP addresses to send or receive data over the network. IoT devices are getting closer to users in their day-to-day activities due to ease of usage and applications-based task performance. This holistic development is raising security concerns, and there is considerable literature in this area. Few IoT-based applications are home automation systems, air pollution monitoring, smart health monitoring, smart traffic management, early flood management, anti-theft and smart agriculture [7]. In these applications, any security concerns can steal sensitive data, launch attacks like blackhole, sinkhole, flooding, or degrade the applications' efficiency, performance, and reliability [8]. Therefore, existing security solutions to address these security issues, particularly edge computing-based intrusion detection systems, are reviewed and discussed in this section.

The Intrusion Detection System (IDS) continuously monitors the incoming data stream from the IoT devices and detect any possible intrusions in the system. There are two major types of IDS: signature-based and anomaly-based IDS. In signature-based, IDS compares the predefined rules, defined as a signature, with the event and reports a threat when there is a match. In comparison, an anomaly-based IDS observes a sequence of events and builds a model for normal behavior. Then the trained model detects the anomalies if there is any change in the pattern. These IDS have their own advantages and limitations. Based on the requirements, the applications can adopt an appropriate IDS. Passban is an intelligent IDS that benefits from edge computing and analyses incoming data on the edge gateways to detect attacks. This system detects various types of malicious traffic like port scanning, HTTP, brute force, and flooding with minimal false-positive rates [9]. IMPACT is a lightweight ML-based IDS to detect impersonation attacks using auto-encoder and feature extraction. This system uses a Support Vector Machine (SVM) method for feature reduction using gradient descent on resource-constrained devices. This step provides efficient training time in identifying the attacks. This ML based-IDS has set a new benchmark on resource-constrained devices for feature reduction [10].

Fair resource-allocation-based IDS secures IoT devices from malware, data hacking, unauthorized access and other system vulnerabilities. Resources are allocated recursively for each node in the edge network. If any edge nodes incur more resource requirements, then an alarm is raised to indicate intrusion in the network. Dominant resource fairness algorithms are used for distributed resources [11]. This approach can restrict the users from demanding more resources and ignore the resources and abilities criteria.

Packet-based IDS employs TCP packets and extracts multiple features of data packets to analyze the intrusions using the Bayesian model. In this approach, data processing is crucial in converting network traffic data into a document, thus making this method the best suitable for applications with multiple data formats and data origins [12]. Similarly, there is another smart data-based IDS using an artificial immune system. Here the cloud server makes a cluster of edge nodes to detect the attacks and generate an alarm in case of intrusions. The smart data concept is an efficient, lightweight method that assists in identifying any silent attacks as well [13]. IDS based on data processing provides critical safety and security to the system since it can alter the system in case of intrusions while sourcing the data from the IoT devices.

Sample selected Extreme Learning Machine (ELM)-based IDS, adaptive ML-based IDS, semi-supervised based, online sequential ELM are all ML-based approaches used to identify intrusions in IoT devices [14]–[17]. In these systems, the edge nodes main functionality is to collect data, process, store and provide services. During training the model with labelled data, the ML algorithms classify attacks accurately and send them to the cloud for decision-making. The cloud server performs the global management of the applications and controls the system during anomalies. There is a proven instance of edge nodes detecting the attack at a 25% faster rate using ML algorithms. Apart from these, there are Deep Neural Network (DNN)-based IDS. The deep neural network is a subset of ML, and it is widely used in the cyber security system to identify unknown data patterns. It uses multiple layers of transformations to find higher-level features. Like ML algorithms, even DNN uses feature selection, training, and testing [18], [19]. Based on the overall literature review considering the advantage and ease of deploying IDS using ML algorithms, the following section includes a comparative study of ML classifiers.

## III. Comparative Analysis

In ML-based IDS, feature selection is an important step to improve the efficiency of the input data from IoT devices. It is the process of selecting required data by excluding irrelevant, redundant, and noisy features, thus bringing a noticeable effect for intrusion detection [20]. The proposed work considers the following six algorithms for the comparative study: Linear Discriminant Analysis (LDA), Logistic Regression (LR), Support Vector Machine (SVM), Decision Tree Classifier (DTC), Random Forest Classifier (RFC), and Multi-Layer Perceptron (MLP). The algorithms first train the dataset so that the classifiers will not be biased towards the most frequent records. Later it is tested to find the accuracy in the prediction. The following section includes brief definitions of these algorithms before analysis.

### A. Linear Discriminant Analysis

LDA is a linear and binary supervised algorithm that distinguishes the data between two classes and assigns the label to known and unknown data. There are three discrimination rules in LDA. This study uses Fisher's linear discriminant rule, which maximizes the ratio between class and within-in class groups to predict the attacks [21]. This data separability between the classes simplifies the dimensions and significantly support in identifying the intrusions.

*B. Logistic Regression*

LR and LDA are similar discriminant analysis methods, but LR is suitable for smaller sample size data. If the sample size during analysis is considered smaller, then the prediction can be more accurate. LR considers sigmoid logistic functions and interprets the probability of the dependent variables [22]. The precise prediction for a smaller sample size with no assumption is the main advantage of this approach.

*C. Support Vector Machine*

SVM is an infinite-dimensional space algorithm suitable for classification, regression or any outliner detection. Unlike prevision methods, SVM creates a line or hyperplane to distinguish the data between the classes. It uses kernel tricks to transform the data and find the optimal boundary between the possible outputs [21]. SVM solves complex optimization problems analytically and returns optimal hyperplane parameters.

*D. Decision Tree Classifier*

DTC is a predictive modelling classifier for continuous values. It is simple and interprets efficiently with little data preparation. Compared to the above methods, there is no data normalization since the trees can handle qualitative predictors with no dummy variables [23]. The rich set of rules available in DTC supports integrating it in real-time applications.

*E. Random Forest Classifier*

RFC is an ensemble learning method for classification that operates by constructing many decision trees during training and selects the appropriate decisions from the output. Compared to DTC, RFC influences the data characteristics and provides better predictions [24]. The automatic balance between huge data is the main advantage of RFC.

*F. Multi-Layer Perceptron*

MLP is a pattern recognition classifier sometimes referred to as a feedforward network. It performs the backpropagation technique for training the data and distinguishes the data that is not linearly separable. It has activation functions that create a linear function between input, output and the hidden layer and support the data interpretations [25]. MLP can predict the projection given a new situation.

IV. EXPERIMENTAL SETUP

The Jupyter on Pycharm tool has been used to perform the comparative analysis using the NSL-KDD dataset, a benchmark for modern-day traffic. It includes train and test data with the most challenging internet traffic records. It contains four different classes of attacks like Denial-of-Service, Probe, User-to-Root, and Remote-to-Local. This dataset includes a total of 43 features per record, out of which 41 features are traffic input, and the last two labels indicate normal or attack data and scores indicating the severity of the attack [26]. The comparative analysis uses a training time and confusion matrix to evaluate the efficiency of the algorithms.

The confusion matrix is a summary of the prediction results of the classification problem. Table 1 shows the confusion matrix that includes the following values: TP (True Positive) represents the intrusions correctly classified; FN (False Negative) represents intrusions misclassified; FP (False Positive) represents non-intrusions misclassified; and TN (True Negative) is the correct prediction of non-intrusion.

TABLE 1. Confusion Matrix

|  |  | True class | |
|---|---|---|---|
|  |  | Positive | Negative |
| Predicted class | Positive | TP | FP |
|  | Negative | FN | TN |

*A. Training Time*

The total time required to train the model is training time. Table 2. illustrates the training time of the considered methods. The results show that SVM has a higher training time of 193.27 seconds due to kernel techniques.

Table 2. Training Time

| ML Algorithms | Training Time (seconds) |
|---|---|
| LDA | 2.54 |
| LR | 2.77 |
| SVM | 193.27 |
| DTC | 1.73 |
| RFC | 12.61 |
| MLP | 1.20 |

*B. Training and Testing Accuracy*

The number of test cases correctly identifies the intrusions is the accuracy of the model. The accuracy obtained by applying the model on the training dataset is called training accuracy. To verify the accuracy obtained, the model is tested with the unknown test dataset, and it is called testing accuracy. Equation (1) shows the evaluation of these accuracies. Fig. 1 shows the training and testing accuracy of the algorithms. DTC has significantly lesser test accuracy than the other methods due to the analytical comparison of the obtained results, and any slight change can drastically change the results.

$$\frac{TP + TN}{TP + TN + FP + FN} \quad (1)$$

*C. Other Performance Evaluation Parameters*

In addition, the following performance evaluation parameters are evaluated using a confusion matrix.

**Precision**: The ratio of all positive labelled instances to the notion of the positive cases that the model accurately recognizes, is known as precision. Equation (2) shows the evaluation of precision.

$$\frac{TP}{TP + FP} \quad (2)$$

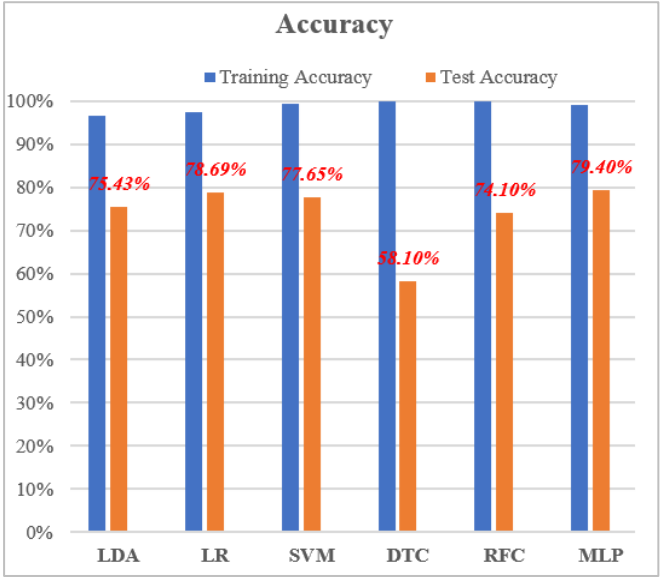

Fig. 1. Training and Testing Accuracy comparison.

**Recall**: The proposition of the positive cases that the model correctly recognizes is called recall. Equation (3) shows the evaluation of recall.

$$\frac{TP}{TP + FN} \quad (3)$$

**F-measure:** A harmonic mean of precision and recall is used to calculate the F-measure and to test accuracy. Equation (4) shows the evaluation of the F-measure.

$$\frac{2X \; Precision \; X \; Recall}{Precision + Recall} \quad (4)$$

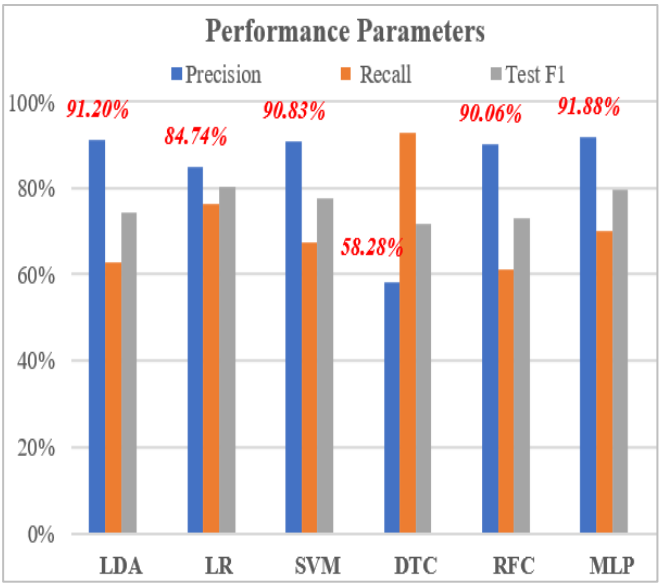

Fig. 2. Performance evaluation comparison.

Fig. 2 shows the overall results obtained from the confusion matrix. The results show that DTC has comparatively less precision, although it has a better recall of 92.8%. And all the other methods have an appropriate harmonic mean balance showing an F-measure of above 70%.

## V. DISCUSSION AND FINDINGS

The comparative analysis discussed shows that MLP and DTC have a better training time. MLP is an efficient classifier based on network patterns. MLP has dependencies between input and output; the output significantly predicts it if the input data has an intrusion data stream. Moreover, there is no normalization in data preprocessing with no scaling in DTC, so the training time is comparatively less. But when confirming the attack, the accuracy is not good enough. It is mainly because a small change in input data causes a larger change in the tree structure, which causes instability. Therefore, it shows a significant difference in the accuracy of the algorithms.

The algorithms considered in this paper have efficient training accuracy of 99%. These algorithms are capable of mapping inputs to the desired outputs efficiently for a given set of predictions. But for edge-based IoT applications and networks, providing the prediction at a faster rate is most important. Apart from SVM, all the other algorithms have better training time, mainly due to many parameters considered during SVM classification. Moreover, to identify the one best algorithm for edge-based IoT applications, these algorithms are further evaluated. The test precision in Fig. 2 indicates that DTC has the least prediction due to the absence of normalization. Finally, LR and MLP have better recall and F-measure, as illustrated in Fig. 2, confirming the balance between recall and precision. However, considering faster training time for predicting attack as a requirement for edge-based IoT network, MLP can be an appropriate algorithm with a training time of 1.2 s, accuracy 99% and F-measure of 79.46%.

Apart from security issues, there are also other areas of improvement for edge-enabled IoT networks. Due to the resource limitations of the edge IoT devices, the efficient resource allocation and intrusion detection system is a challenging research area. In addition, the benefits of using machine learning-based intrusion detection include the reduction in load for edge nodes, the possibility of deployment of fine-grained security mechanisms, and the reduction of interaction among nodes for security purposes. Moreover, federated learning techniques are emerging further to enhance intrusion detection systems [27].

## VI. CONCLUSION AND FUTURE WORK

As IoT devices are increasing in day-to-day activities, ensuring the security of these devices has become more prevalent. Although IDS is a mature research area, it still requires intense effort towards securing IoT applications through edge computing. Unlike conventional security methods, IDS deployed in edge computing need significant support to identify intrusions efficiently with minimal resources and computational capabilities. In this paper, we performed a comparative study of ML algorithms for edge-based IoT applications. ML algorithms can improve efficiency over time and support computationally intensive edge-based applications without overloading them. From the comparison, we observed that MLP has better training time. MLP has dependencies between input and output and is an efficient classifier for edge-based IDS. And it provides better prediction with minimal time and makes it suitable for edge-based IoT networks. In future, we will further enhance this algorithm to develop a secure edge-enabled IoT network.

ACKNOWLEDGMENT

"This research work is supported by Taylor's University, Malaysia."

REFERENCES

[1] I. Lee, "Internet of Things (IoT) cybersecurity: Literature review and iot cyber risk management," *Futur. Internet*, vol. 12, no. 9, 2020, doi: 10.3390/FI12090157.

[2] D. Halperin *et al.*, "Pacemakers and implantable cardiac defibrillators: Software radio attacks and zero-power defenses," *Proc. - IEEE Symp. Secur. Priv.*, pp. 129–142, 2008, doi: 10.1109/SP.2008.31.

[3] L. Tawalbeh, F. Muheidat, M. Tawalbeh, and M. Quwaider, "applied sciences IoT Privacy and Security : Challenges and Solutions," *Mdpi*, pp. 1–17, 2020.

[4] V. Hassija, V. Chamola, V. Saxena, D. Jain, P. Goyal, and B. Sikdar, "A Survey on IoT Security: Application Areas, Security Threats, and Solution Architectures," *IEEE Access*, vol. 7, pp. 82721–82743, 2019, doi: 10.1109/ACCESS.2019.2924045.

[5] J. Ni, K. Zhang, X. Lin, and X. Shen, "Securing Fog Computing for Internet of Things Applications: Challenges and Solutions," *IEEE Commun. Surv. Tutorials*, vol. 20, no. 1, pp. 601–628, 2018, doi: 10.1109/COMST.2017.2762345.

[6] R. Gupta, S. Tanwar, S. Tyagi, and N. Kumar, "Machine Learning Models for Secure Data Analytics: A taxonomy and threat model," *Comput. Commun.*, vol. 153, no. November 2019, pp. 406–440, Mar. 2020, doi: 10.1016/j.comcom.2020.02.008.

[7] S. Pundir, M. Wazid, D. P. Singh, A. K. Das, J. J. P. C. Rodrigues, and Y. Park, "Intrusion Detection Protocols in Wireless Sensor Networks Integrated to Internet of Things Deployment: Survey and Future Challenges," *IEEE Access*, vol. 8, pp. 3343–3363, 2020, doi: 10.1109/ACCESS.2019.2962829.

[8] S. M. Muzammal, R. K. Murugesan, and N. Z. Jhanjhi, "A Comprehensive Review on Secure Routing in Internet of Things: Mitigation Methods and Trust-based Approaches," *IEEE Internet Things J.*, pp. 1–1, 2020, doi: 10.1109/JIOT.2020.3031162.

[9] M. Eskandari, Z. H. Janjua, M. Vecchio, and F. Antonelli, "Passban IDS: An Intelligent Anomaly-Based Intrusion Detection System for IoT Edge Devices," *IEEE Internet Things J.*, vol. 7, no. 8, pp. 6882–6897, 2020, doi: 10.1109/JIOT.2020.2970501.

[10] S. J. Lee *et al.*, "IMPACT: Impersonation Attack Detection via Edge Computing Using Deep Autoencoder and Feature Abstraction," *IEEE Access*, vol. 8, pp. 65520–65529, 2020, doi: 10.1109/ACCESS.2020.2985089.

[11] F. Lin, Y. Zhou, X. An, I. You, and K. K. R. Choo, "Fair Resource Allocation in an Intrusion-Detection System for Edge Computing: Ensuring the Security of Internet of Things Devices," *IEEE Consum. Electron. Mag.*, vol. 7, no. 6, pp. 45–50, 2018, doi: 10.1109/MCE.2018.2851723.

[12] X. Cao, Y. Fu, and B. Chen, "Packet-based intrusion detection using Bayesian topic models in mobile edge computing," *Secur. Commun. Networks*, vol. 2020, 2020, doi: 10.1155/2020/8860418.

[13] F. Hosseinpour, P. Vahdani Amoli, J. Plosila, T. Hämäläinen, and H. Tenhunen, "An Intrusion Detection System for Fog Computing and IoT based Logistic Systems using a Smart Data Approach," *Int. J. Digit. Content Technol. its Appl.*, vol. 10, no. 5, pp. 34–46, 2016.

[14] X. An, X. Zhou, X. Lü, F. Lin, and L. Yang, "Sample selected extreme learning machine based intrusion detection in fog computing and MEC," *Wirel. Commun. Mob. Comput.*, vol. 2018, 2018, doi: 10.1155/2018/7472095.

[15] Y. Wang, W. Meng, W. Li, Z. Liu, Y. Liu, and H. Xue, "Adaptive machine learning-based alarm reduction via edge computing for distributed intrusion detection systems," *Concurr. Comput.*, vol. 31, no. 19, pp. 1–12, 2019, doi: 10.1002/cpe.5101.

[16] S. Xu, Y. Qian, and R. Q. Hu, "A Semi-Supervised Learning Approach for Network Anomaly Detection in Fog Computing," *IEEE Int. Conf. Commun.*, vol. 2019-May, 2019, doi: 10.1109/ICC.2019.8761459.

[17] S. Prabavathy, K. Sundarakantham, and S. M. Shalinie, "Design of cognitive fog computing for intrusion detection in Internet of Things," *J. Commun. Networks*, vol. 20, no. 3, pp. 291–298, Jun. 2018, doi: 10.1109/JCN.2018.000041.

[18] Z. Liu, X. Yin, and Y. Hu, "CPSS LR-DDoS Detection and Defense in Edge Computing Utilizing DCNN Q-Learning," *IEEE Access*, vol. 8, no. 3, pp. 42120–42130, 2020, doi: 10.1109/ACCESS.2020.2976706.

[19] K. Sadaf and J. Sultana, "Intrusion detection based on autoencoder and isolation forest in fog computing," *IEEE Access*, vol. 8, pp. 167059–167068, 2020, doi: 10.1109/ACCESS.2020.3022855.

[20] H. Chae, B. Jo, S. Choi, and T. Park, "Feature Selection for Intrusion Detection using NSL-KDD," *Recent Adv. Comput. Sci. 20132*, pp. 184–187, 2013.

[21] A. Dellacasa Bellinegni *et al.*, "NLR, MLP, SVM, and LDA: A comparative analysis on EMG data from people with trans-radial amputation," *J. Neuroeng. Rehabil.*, vol. 14, no. 1, pp. 1–17, 2017, doi: 10.1186/s12984-017-0290-6.

[22] P. Subramaniam and M. J. Kaur, "Review of Security in Mobile Edge Computing with Deep Learning," in *2019 Advances in Science and Engineering Technology International Conferences (ASET)*, Mar. 2019, no. June, pp. 1–5, doi: 10.1109/ICASET.2019.8714349.

[23] F. H. Botes, L. Leenen, and R. De La Harpe, "Ant colony induced decision trees for intrusion detection," *Eur. Conf. Inf. Warf. Secur. ECCWS*, no. June, pp. 53–62, 2017.

[24] W. Pang, H. Jiang, and S. Li, "Sparse Contribution Feature Selection and Classifiers Optimized by Concave-Convex Variation for HCC Image Recognition," *Biomed Res. Int.*, vol. 2017, 2017, doi: 10.1155/2017/9718386.

[25] B. S. Khater, A. W. B. A. Wahab, M. Y. I. Bin Idris, M. A. Hussain, and A. A. Ibrahim, "A lightweight perceptron-based intrusion detection system for fog computing," *Appl. Sci.*, vol. 9, no. 1, 2019, doi: 10.3390/app9010178.

[26] C. I. for Cybersecurity, "NSL-KDD | Datasets | Research | Canadian Institute for Cybersecurity | UNB." https://www.unb.ca/cic/datasets/nsl.html (accessed Apr. 20, 2020).

[27] K. Bonawitz *et al.*, "Towards federated learning at scale: System

design," *arXiv*, 2019.

[28] S. A. Rahman, H. Tout, C. Talhi, and A. Mourad, "Internet of Things intrusion Detection: Centralized, On-Device, or Federated Learning?," *IEEE Netw.*, vol. 34, no. 6, pp. 310–317, Nov. 2020, doi: 10.1109/MNET.011.2000286.